\documentclass[traditabstract,a4]{aa}  
\usepackage{graphicx,natbib,amsmath,amssymb,psfig}
\def\hmpc{~h^{-1}{\rm Mpc}}
\def\hcmpcc{~h^{3}{\rm Mpc}^{-3}}
\def\hinvm{h^{-1}{\rm M_\odot}}

\def\deg2{{\rm~deg}^2}
\def\degm2{\rm~deg^{-2}}
\def\arc2{{\rm~arcmin}^2}
\begin{document}
   \title{The WIRCam Deep Survey II: Mass Selected Clustering\thanks{Based on data obtained with the European Southern Observatory Very Large
Telescope, Paranal, Chile, under Large Programs 070.A-9007, 175.A-0839, and 177.A-0837. }}

   \author{R. M. Bielby\inst{1}, V. Gonzalez-Perez\inst{1,2}, H. J. McCracken\inst{3}, O. Ilbert\inst{4}, E. Daddi\inst{5}, O. Le F{\`e}vre\inst{4}, P. Hudelot\inst{3}, J.-P. Kneib\inst{4}, Y. Mellier\inst{3}, C. Willott\inst{6}\fnmsep}

   \institute{
Dept. of Physics, Durham University, South Road, Durham, DH1 3LE, UK
    \and
Centre de Physique des Particules de Marseille, Aix-Marseille Universit\'e,  CNRS/IN2P3, Marseille, France
	\and	
Institut d'Astrophysique de Paris, UMR7095 CNRS, Universit\'e  Pierre et Marie Curie, 98 bis Boulevard Arago, 75014 Paris, France
    \and
 Aix Marseille Universit\'{e}, CNRS, LAM - Laboratoire d'Astrophysique de Marseille, 38 rue F. Joliot-Curie, F-13388, Marseille, France    \and
Service d'Astrophysique, CEA/Saclay, 91191 Gif-sur-Yvette, France
    \and
Herzberg Institute of Astrophysics, National Research Council, 5071 West Saanich Road, Victoria, BC V9E 2E7, Canada\\
\email{rmbielby@gmail.com}
}

   \date{}

  \abstract{We present an analysis of the clustering of galaxies from $z\approx2$ to the present day using the WIRCam Deep Survey (WIRDS). WIRDS combines deep optical data from the CFHTLS Deep fields with its' own deep near-infrared data, providing a photometric data-set over an effective area of 2.4 deg$^2$, from which accurate photometric redshifts and stellar masses can be estimated. We use the data to calculate the angular correlation function for galaxy samples split by star-formation activity, stellar mass and redshift. Using WIRDS with its' large total area and multiple fields gives a low cosmic variance contribution to the error, which we estimate to be less than $\sim2.8\%$. Based on power-law fits, we estimate the real-space clustering for each sample, determining clustering lengths and power-law slopes. For galaxies selected by constant mass, we find that the clustering scale shows no evolution up to $z\approx2$. Splitting the galaxy sample by mass, we see a consistent trend for higher mass galaxies to have larger clustering scales at all redshifts considered. We use our results to test the {\sc galform} semi-analytical model of galaxy formation and evolution. Our results are well matched by the model galaxies for both the redshift evolution and the mass dependence of the galaxy clustering. We split the galaxy population into passive and star-forming populations based on rest-frame dust-corrected NUV-$r$ colours. We find that the passive galaxy populations show a significantly larger clustering scale at all redshifts than the star-forming population below masses of $M_\star\sim10^{11}~\hinvm$, showing that even at $z\approx2$ passive galaxies exist in denser environments than the bulk of the star-forming galaxy population. For star-forming galaxies with stellar masses of $M_\star\gtrsim10^{11}~\hinvm$, we find a clustering strength of $\sim8\hmpc$ across all redshifts, comparable to the measurements for the passive population. Additionally, for star-forming galaxies we see that clustering strength increases for higher stellar mass systems, however little sign of a mass dependence in passive galaxies is observed over the range in stellar mass that is probed. Comparing our results to the model galaxy population produced by {\sc galform}, we find good agreement between the model predictions and the observed clustering. Finally, we investigate the connection between galaxy stellar mass and dark matter halo mass, showing a clear correlation between the two in both the WIRDS data and the {\sc galform} predictions.}

   \keywords{galaxies: evolution --- galaxies: high-redshift --- (cosmology:) large scale structure of the universe}
\authorrunning{R. Bielby et al}
\titlerunning{The WIRCam Deep Survey II}
   \maketitle
%

\section{Introduction}

\setcounter{footnote}{0}

The current consensus suggests that star formation in the Universe reached its peak within the redshift range $1<z<2$, whilst $\sim 50\%-70\%$ of mass assembly took place in the redshift range $1<z<3$ \citep{1997ApJ...486L..11C,2003ApJ...587...25D,2007A&A...476..137A, 2007A&A...474..443P, 2007ApJ...660L..43N}. It is evident that there is a substantial population of massive galaxies at $z>1$ and this has presented significant complications for hierarchical structure formation models. Additional observations of galaxies at $z>1$ are of paramount importance to advance our understanding of galaxy formation and evolution as a whole. 


At redshifts of $z>1$, identifiable spectral features begin to move out of the optical wavelength range and near-infrared observations become essential. The role of environment and large-scale structure at these redshifts is largely unexplored \citep{2009arXiv0906.4662R}. In addition to making it possible to select galaxies in this important redshift range, near-infrared galaxy samples offer several advantages compared to purely optical selections (see for example \citealt{1994ApJ...434..114C}). As $k-$corrections in $K_s$ band are insensitive to galaxy type over a wide redshift range, near-infrared-selected samples provide a fairly unbiased census of galaxy populations at high redshifts (providing that the extinction is not too high, as in the case of some submillimeter galaxies). Such samples represent the ideal input catalogues from which to extract targets for spectroscopic surveys as well as for determining accurate photometric redshifts.

An important application and aim for achieving deeper NIR data is to trace galaxy formation and evolution through the period of high star-formation at $1<z<2$. Key in this is understanding the evolution of the spatial clustering of galaxies to the present day. From such analysis of the spatial distribution of galaxies, one can derive the mass of the dark matter halos in which they reside. It is then crucial to quantify the links that may exist between the dark matter halos and the attributes of the galaxies themselves, in particular the stellar-mass, galaxy type, luminosity and star-formation rate. In addition, given an estimate of the mass of the dark matter haloes hosting a given galaxy population (and assuming a suitable model for halo evolution) one can identify the present-day descendants of these galaxies, as has been done for Lyman break galaxies (LBGs) at $z\sim3$ \citep[e.g. ][]{2006ApJ...652..994C,2008ApJ...679.1192C,2011MNRAS.414....2B}. A few studies have attempted this for passive galaxies at $z\sim2$, but small fields of view have made these studies somewhat sensitive to the effects of cosmic variance.

Analysing the spatial distribution of galaxies and understanding the relationship between this distribution and galaxy properties is a key element in the study of galaxy formation theory. Studies of the relationship between clustering and morphology \citep[e.g.][]{1976ApJ...208...13D,1993MNRAS.265...21I,1995ApJ...442..457L,2002MNRAS.332..827N,2012A&A...542A...5C} have for a long time registered the stronger clustering tendencies of early-type over late-type galaxies. Most recently, \citet{2011MNRAS.412..825D} used HST imaging of the COSMOS field to determine the morphological types of galaxies in zCOSMOS and confirmed the stronger clustering of early type galaxies over late type galaxies to $z\approx1$. Extensive work has also been performed on the relationship between galaxy luminosity and clustering, with greater luminosity correlating with greater clustering as observed at lower redshifts \citep[e.g.][]{1987MNRAS.229..621P,1993MNRAS.263..191H,2001MNRAS.328...64N} and extending to $z\sim1$ \citep{2006A&A...451..409P,2006A&A...452..387M,2008ApJ...672..153C}.

Extending these studies to $1<z<2$ has been made possible in the last decade through the increasing availability of deep NIR imaging data as well as extensive spectroscopic redshift surveys \citep{2005A&A...439..877L,2006ApJ...644..671C}. Focusing on the NIR imaging, the relative difficulty in obtaining sufficiently deep data has led many studies to focus on tracer populations identified using colour selection techniques. Perhaps the most successful of these has been the $BzK$ selection \citep{daddi04}, which facilitates the selection of galaxies at $z\gtrsim1.4$ and the approximate separation of these into passive (p$BzK$) and star-forming (s$BzK$) populations. Several surveys have applied the $BzK$ selection techniques to large samples of near-infrared selected galaxies. In one of the widest surveys to date, \cite{2006ApJ...638...72K} constructed $K$-band selected samples over a $\sim 920$ arcmin$^2$ field reaching $K\approx 20.8$ and attaining $K\approx 21.8$ over a 320 arcmin$^2$ sub-field. The exploration of a field of this size made it possible to measure the clustering properties of star-forming and passive galaxy samples and to establish that passive galaxies in this redshift range are substantially more strongly clustered than star-forming ones, indicating that a galaxy-type - density relation reminiscent of the local morphology-density relation must be already in place at $z>1.4$. Subsequent studies using the MUSYC \citep{blanc08}, UKIDSS \citep{2008MNRAS.391.1301H} and COSMOS \citep{2010ApJ...708..202M} surveys have confirmed these results, whilst also establishing the continuation of the luminosity dependence of galaxy clustering beyond $z>1$ \citep[e.g. ][]{2007ApJ...660...72H}.

Complimentary studies at $z>1$ have also been performed using a variety of colour selection techniques.  For example, galaxies selected as Extremely Red Objects (EROs, isolated using a cut in the $R-K$ colour) were found to be highly clustered and indicated the existence of a $z>1$ elliptical galaxy population \citet{2000A&A...361..535D,2002MNRAS.337.1282R,2005ApJ...621...41B}. Similarly, Distant Red Galaxies (DRGs, selected via $J-K$ colours) have been shown to be highly clustered galaxies at $1<z<3$ \citep{2006A&A...453..507G,2007MNRAS.376L..20F,2008ApJ...685L...1Q,2011MNRAS.410..241K}, whilst optically selected galaxies have also played their part \citep{2005ApJ...619..697A}.

However, given the increasing availabilty of multi-band photometry through optical and NIR wavebands, more complex selections of galaxy populations are becoming feasible and reliable via photometric template fitting. \citet{2010MNRAS.409..184P} used the MUSYC survey to evaluate the clustering of galaxies to $z\approx1.5$, showing a mild dependence on sample luminosity out to this distance. \citet{2010MNRAS.407.1212H} went further with the UKIDDS data, analysing the clustering of the passive and star-forming galaxy populations to $z\approx3$ and confirming the stronger clustering of the passive population over the star-forming to $z\approx1.5$, whilst reporting little dependence of clustering on $K$ band luminosity. Interestingly they find that the clustering of star-forming galaxies increases with redshift and reaches equivalent strengths to the clustering of passive galaxies at $z>2$.

It has only been relatively recently with the advent of the deep NIR imaging surveys and extensive spectroscopic surveys that the relationship of clustering to stellar mass have begun to be deeply investigated. For example, first taking the spectroscopic based work, \citet{2006MNRAS.368...21L} showed the measured the mass dependence of galaxy clustering at $z<0.3$ using SDSS data, showing an increase in clustering with mass, which became more pronounced above $M^*$. Moving to higher redshifts, \citet{2008A&A...478..299M} and \citet{2009A&A...505..463M} measured the clustering of mass-selected samples in the VVDS-Deep and zCOSMOS Surveys respectively, finding a clustering mass dependence in their results at redshifts up to $z\approx1.2$.

Returning to photometric data, \citet{2010MNRAS.406..147F} used Palomar Observatory Wide-field Infrared Survey (POWIS) to measure the mass dependency of galaxy clustering to $z=2$, over a combined field of view of 1.16 deg$^2$ and with $K$ depths of $\approx23.5$. Taking the full galaxy population, they found an increase in galaxy clustering with galaxy stellar mass across a range of redshifts, whilst also noting an increase in the clustering strength with redshift for samples of the same mass range. Similarly, \citet{2011ApJ...728...46W} measured the clustering of galaxies as a function of mass in the 0.4 deg$^2$ of the NEWFIRM (NOAO Extremely Wide-Field Infrared Imager) Medium Band Survey (NMBS). Again they point to a strong dependence of galaxy clustering on galaxy stellar mass. In terms of any stellar mass/halo mass relation, \citet{2011ApJ...728...46W} see little evidence of any redshift dependence in the relationship between stellar mass and halo mass over the range $1<z<2$, but see evidence for a change in the relation when comparing to results at $z<1$ from other surveys.

In this paper, we present a study of the mass, type and redshift dependence of galaxy clustering in the WIRCam Deep Survey (WIRDS). The paper is organized as follows: Section~\ref{sec:obs-data} briefly describes the WIRDS data used here. Following this, in Section~\ref{sec:clustering} we present the clustering analysis if galaxies in the WIRDS fields as a function of type, mass and redshift over the range $0<z<2$. Section~\ref{sec:conclusions} provides a summary and our conclusions.

Throughout this paper, all magnitudes are given in the AB system unless stated otherwise. Where relevant, we assume a $\Lambda$CDM cosmology given by $\Omega_m=0.25$ and $\Omega_\Lambda=0.75$.

\section{Data and Simulations}
\label{sec:obs-data}

The work presented here is based on data from the deep fields of the Canada-France-Hawai'i Telescope Legacy Survey (CFHTLS). The CFHTLS Deep incorporates four $1\times1$ deg$^2$ survey fields (designated D1 to D4) spread across a broad range in R.A. and Declination, the four field centre co-ordinates being: D1 02:25:59, $-$04:29:40; D2 10:00:28, $+$02:12:30; D3 14:19:27, $+$52:40:56; and D4 22:15:31, $-$17:43:56. The CFHTLS produced deep optical imaging data within these fields, whilst the WIRDS survey has added deep NIR imaging data and it is the combination of these two datasets that we use here and that are described in more detail in the two sections below. In this section, we also give an overview of the {\sc galform} semi-analytical galaxy formation model, the predictions of which we will confront with our results.

A key benefit of the WIRDS data is the combination of 4 distinct large deep fields, which act to reduce the impact of cosmic variance. Based on the cosmic variance cookbook ({\sc getcv}) of \citet{2011ApJ...731..113M}, we estimate the contribution to the errors on our clustering measurements to be $\lesssim2.8\%$.

\subsection{Optical data}

In this work, we use the CFHTLS T0006 optical data to provide 5-band optical photometry of the galaxy population. This incorporates imaging taken with the MegaCAM imager in the Canada-France-Hawai'i Telescope (CFHT) using the $ugriz$ MegaCAM filters. The CFHTLS provides stacked images based on either a 25\% or 85\% cut in terms of image quality (i.e. seeing). In this case, we use catalogues based on the 85\%-cut stacks. These catalogues have point-like (i.e. star/bulge) 80\% completeness depths of $u\approx26.5$, $g\approx26.0$, $r\approx25.5$, $i\approx25.1$ and $z\approx24.9$, whist the image quality is consistently $\approx0.7-0.8''$ across all four fields. This consistency is one of the key benefits of using the CFHTLS data, in that all the data is homogeneous in terms of seeing, depth and filter/telescope properties. A full and thorough description and characterisation is presented by \citet{CFHTLS_T0006} and we refer the reader to this document for any further information.

\subsection{Infrared data}

WIRDS is a deep infrared imaging survey of the CFHTLS Deep fields, providing infrared data to complement the CFHTLS optical data obtained with MegaCAM. The WIRDS imaging was taken with the WIRCam detector \citep{puget04} on the CFHT and a detailed discussion of the observations and data reduction of the WIRDS data is provided by \citet{2012A&A...545A..23B} and \citet{2010ApJ...708..202M}.

The data used in this paper were taken in a series of observing run from 2005-2007 and were made in co-ordination with the COSMOS consortium. Observations were conducted using three filters: $J$, $H$ and $K_s$. Transmission plots of the WIRCam $J$, $H$ and $K_s$ filters are available from CFHT\footnote{http://www.cfht.hawaii.edu/Instruments/Filters/wircam.html}. The integration times for all $J$, $H$ and $K_s$ band exposures was 45s, 15s and 20s respectively.

The observations were carried out in queue scheduled mode at the CFHT. Image quality constraints of $0.55\arcsec<IQ<0.65\arcsec$ were requested and the observations were micro-dithered using the standard WIRCam micro-dither pattern consisting of 2$\times$2 dither patter with offsets between consecutive dithers of 0.5 pixels. Due to the WIRCam pixel scale of $0.3\arcsec$/pixel, this micro-dithering is required in order to produce well sampled images under our seeing constraints (and to allow matching with the CFHTLS pixel-scale of $0.186\arcsec$/pixel). A further large-scale dithering pattern was applied to the observations to avoid gaps in the coverage due to the gaps between adjacent CCDs.

The photometric catalogues on which this work is based are available at the CADC archive (http://cadcwww.dao.nrc.ca/cfht/WIRDST0002.html).

\subsection{Photometric galaxy properties, mass constraints and colour selection}

With the wavelength coverage afforded by the combination of the CFHTLS optical data and the WIRDS NIR data, it is possible to estimate photometric redshifts and stellar masses reliably over a broad redshift range. In particular, the wavelength range afforded by this collection of filters presents the possibility of the $4000\AA$ break being detectable up to $z\sim4$.

We used the Le Phare\footnote{http://www.cfht.hawaii.edu/$\sim$arnouts/LEPHARE/cfht\_lephare} code \citep{arnouts02,ilbert06} to determine photometric redshifts and galaxy properties with a $\chi^2$ template-fitting method. The photo-z were estimated using the median of the probability distribution function (PDFz) rather than the minimum of the $\chi^2$ distribution. The results of the photometric redshift estimation are presented in \citet{2012A&A...545A..23B}, with a full comparison to spectroscopic datasets. Below we provide an overview and derived accuracies of the photometric redshifts.

We use a number of spectroscopic redshift data-sets to calibrate the photo-z data in our four fields. In the D1 field, we use spectroscopic redshifts from the VVDS Deep \citep{2005A&A...439..845L} and Ultra-Deep \citep{2012A&A...539A..31C,2013arXiv1307.6518L} spectroscopic samples. The VVDS Deep sample is available publicly and consists of 8,981 spectroscopically observed objects over an area of $0.5\deg2$ in the CFHTLS D1 field. It consists of a magnitude limited sample with a limit of $I\leq24$ and samples a redshift range of $0\leq z\leq5$. The Ultra-Deep sample consists of $\sim1500$ spectra over an area of $\approx0.14\deg2$ and covers a magnitude range of $22.5\leq i\leq24.75$. Both of the VVDS spectroscopic catalogues attribute each object a flag based on the identification. These range from 1 to 4 with 1 being most unreliable and 4 being most reliable. In addition a flag 9 is given to objects identified based on a single emission line. Using the VVDS Deep data we find an outlier rate of $\eta=3.7\%$ and $\sigma_{\Delta z/(1+z)}=0.025$, with a median magnitude of $i^\ast_{median}=24.0$, whilst using the UltraDeep data we find an outlier rate of $\eta=4.2\%$ and $\sigma_{\Delta z/(1+z)}=0.030$, with a median magnitude of $i^\ast_{median}=23.7$.

In the D2/COSMOS field we make use of the zCOSMOS 10k data \citep{2009ApJS..184..218L}, which constituted the ESO Large Proposal LP175.A-0839 and provides spectroscopic redshifts based on data acquired using VIMOS on the VLT. We find 3004 objects predominantly in the magnitude range $17.5\leq i\leq22.5$ and over a redshift range up to $z\lesssim1.4$, which are present in our photometric catalogue. From this data (and using only objects with spectroscopic flags of 3 or 4) we estimate an outlier rate of $\eta=1.4\%$ and $\sigma_{\Delta z/(1+z)}=0.023$, based on a sample with median magnitude of $i^\ast_{median}=21.6$.

In D3, we use the DEEP2 DR3 redshift catalogue \citep{2003SPIE.4834..161D,2007ApJ...660L...1D}, which is based on spectroscopic observations using the Deep Imaging Multi-Object Spectrograph (DEIMOS) on Keck II. The catalogue contains 47,700 unique objects, of which 2,977 have a `zquality' flag of $\geq3$ (i.e. are deemed to be reliable redshifts) and are present in our photometric catalogue. This sample predominantly covers a magnitude range of $18<i<24$ with the bulk being below a redshift of $z\lesssim1.6$. Using the DEEP2 data, we estimate an outlier rate of $\eta=3.4\%$ and $\sigma_{\Delta z/(1+z)}=0.027$, based on a sample with median magnitude of $i^\ast_{median}=22.5$ for the WIRDS D3 photometric catalogue.

In the D4 field, we use spectra obtained using the AAOmega instrument on the Anglo-Australian Telescope (AAT) as part of a program to provide optical spectroscopy of X-ray point-sources in the CFHTLS \citep{2010MNRAS.401..294S}. The observations provide redshifts for 1,800 objects in the D4 field, of which 168 are QSOs, 66 are stars and 1,335 are galaxies, all at magnitudes of $i<22.5$ \citep{2010A&A...523A..66B}. In total, 1,090 of the galaxies overlap with our photometric data, most of which are at $z\lesssim0.8$. Based on these, we find an outlier rate of $\eta=2.1\%$ and $\sigma_{\Delta z/(1+z)}=0.021$, with a median magnitude of $i^\ast_{median}=20.0$.

\begin{figure}
\centering
\includegraphics[width=90.mm]{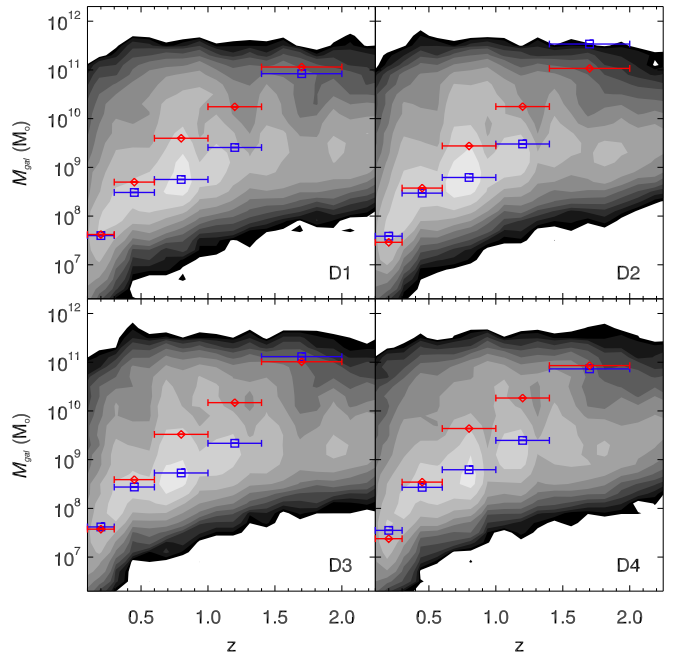}
   \caption{Estimated mass completeness limits of the WIRDS data. The greyscale contours show the distribution of the $i<25$ galaxy population, normalized by the area of each field. Red diamonds with horizontal error bars show the estimated mass completeness limits for consecutive bins in redshift corresponding to a magnitude cut of $i=25$ for passive galaxies. Blue squares with error bars show the same for star-forming galaxies.}
   \label{fig:mass_lim}
\end{figure}

\begin{table}
\caption{Mass completeness limits based on a maximum 30\% of objects at magnitudes of $i>25$.}             
\label{tab:masslimits}
\centering          
\begin{tabular}{l r r }     
\hline\hline       
         & \multicolumn{2}{c}{Median Mass Limit} \\
Redshift & \multicolumn{2}{c}{($\mbox{log}_{10}(M[\hinvm])$)} \\
         & Passive    & Star-forming \\
\hline
$0.3<z\leq0.6$ & 8.6 & 8.5 \\
$0.6<z\leq1.0$ & 9.6 & 8.8 \\
$1.0<z\leq1.4$ & 10.2& 9.4 \\
$1.4<z\leq2.0$ & 11.0& 11.1\\
\hline
\hline                  
\end{tabular}
\end{table}

For the purposes of the clustering analysis, we focus on measuring the clustering of galaxy populations selected using the photometric redshifts and photometrically derived masses, maximising the use of the WIRDS catalogues. We use three overall samples, the passive galaxy sample selected via rest-frame colours, the star-forming galaxy sample selected in the same way and the third incorporating the entire catalogue. The colour selection is the same as that used by \citet{2010ApJ...709..644I} in which galaxies with rest frame, dust de-reddened colours of $NUV-r\geq3.5$ are classed as passive and those with $NUV-r<3.5$ are classed as star-forming. Each of these are then split into redshift and mass slices. 

We first evaluate the mass completeness limits of each of the three samples in the four WIRDS fields. This is done following the method of \citet{2010ApJ...709..644I}, setting our mass completeness limits as the lowest mass at which $<30\%$ of galaxies are fainter than a chosen magnitude limit. We estimate this limit as a function of galaxy redshift and type (i.e. passive, star-forming and both combined) using a magnitude limit of $i=25.0$.

The estimated mass completeness limits are shown in Fig.~\ref{fig:mass_lim} as a function of redshift for each of the fields. The grey-scale contour maps show the galaxy population distribution at $i<25$ normalised by the field areas. Mass completeness limits for the star-forming population are shown by the blue squares, whilst the limits for the passive population are given by the red diamonds. The extent of each redshift-bin is given by the horizontal error bars. Estimated limits across the four fields are broadly consistent based on the imposed $i<25$ limit. In each field there is also a consistent separation between the star-forming and passive galaxy mass-limits, with the star-forming galaxies probing to lower masses at $0.5\lesssim z\lesssim1.4$ given the $i<25$ magnitude limit.

We take the median mass limit of the four fields in five redshift bins and these are given in Table~\ref{tab:masslimits}. Given the low ratio of passive to star-forming galaxies, the limits for the entire sample are equivalent to those of the star-forming sample. Based on these mass constraints and redshift bins we split the galaxy population in each field into samples covering a range in stellar mass. For all three of the passive, star-forming and complete samples, we separate the population into stellar mass bins of $10^{8.6}<M[\hinvm]\leq10^{9.6}$, $10^{9.6}<M[\hinvm]\leq10^{10.6}$, $M[\hinvm]>10^{10.6}$ and $M[\hinvm]>10^{11}$.

\subsection{Simulations}
\label{sec:galform}

We predict the clustering of mass selected galaxies in a $\Lambda$CDM universe using the {\sc galform} semi-analytical galaxy formation code developed by \citet{cole00}, and extended by \citet{2003ApJ...599...38B}, \citet{2005MNRAS.356.1191B}, \citet{2006MNRAS.370..645B}, \citet{lagos10} and \citet{nikos11}. Semi-analytical models use physically motivated recipes and rules to follow the fate of baryons in a universe in which structures grow hierarchically through gravitational instability \citep[see][for an overview of hierarchical galaxy formation models]{baugh06}.

In this paper we focus our attention on the \citet{2006MNRAS.370..645B} model. Some of the key features of this model are (i) a time scale for quiescent star formation that varies with the dynamical time of the disk and which therefore changes significantly with redshift, (ii) bursts of star formation occur due to both galaxy mergers and when disks become dynamically unstable, and (iii) the inclusion of both supernova and active galactic nuclei (AGN) feedback. This feedback is implemented in such a way that AGNs are able to heat the cooling flows in massive haloes, preventing any further star formation in galaxies within such haloes. \citeauthor{2006MNRAS.370..645B} adopt the cosmological parameters of the Millennium Simulation \citep{2005Natur.435..629S}, which are in broad agreement with constraints from measurements of the cosmic microwave background radiation and large scale galaxy clustering \citep[e.g.][]{sanchez09}: $\Omega_{m0}=0.25$, $\Omega_{\Lambda0} = 0.75$, $\Omega_{b0}=0.045$, $\sigma_{8}=0.9$ and $h=0.73$. The \citeauthor{2006MNRAS.370..645B} model parameters were fixed with reference to a subset of the available observations of galaxies, mostly at low redshift. For further details we refer the reader to \citet{2006MNRAS.370..645B}. This model has previously been used for studying the clustering of galaxies at both low \citep[$z\sim0.1$,][]{han09} and high redshifts \citep[$z\sim 1$,][]{eros2}.

We note that GALFORM uses a Kennicutt IMF, whilst the photometric masses derived from the WIRDS data assume a Chabrier IMF. Based on \citet{2010ApJ...709..644I} and \citet{2011MNRAS.414..304G}, we multiply the GALFORM masses by a factor of 1.32 to match the Chabrier based stellar masses of the WIRDS photometric catalogues.

The \citeauthor{2006MNRAS.370..645B} model successfully reproduces the stellar mass function up to $z=4.5$ and the number counts of red galaxies at $z<2$ \citep{2008MNRAS.386.2145A,2009MNRAS.398..497G}. In addition, we have shown in \citet{2012A&A...545A..23B}, that the mass function for all galaxies is well matched between our observations and the {\sc galform} model up to a redshift of at least $z\approx2$.

\section{Galaxy Clustering}
\label{sec:clustering}
\subsection{The Angular Correlation Function}
\label{sec:clust_theory}
The angular correlation function can be measured using a number of estimators. In this paper, we use the Landy-Szalay estimator, which is given by:

\begin{equation}
w(\theta)=\frac{\langle DD\rangle-2\langle DR\rangle+\langle RR\rangle}{\langle RR\rangle}
\end{equation}

\noindent where $\langle DD\rangle$ is the number of galaxy-galaxy pairs, $\langle DR\rangle$ is the number of galaxy-random pairs and $\langle RR\rangle$ is the number of random-random pairs. The Landy-Szalay estimator avoids the issue of linear biases seen in the direct estimator caused by spurious signal between the data and survey window. The random catalogues used to evaluate the estimator are produced using the survey geometry with the optical and NIR masks described above applied. The random galaxy catalogues each contain a total number of random data points equal to $20\times$ the number of galaxies in the `real-data' catalogue with which the correlation function is being calculated.

As discussed, each of our fields measures $\approx0.4-0.8\deg2$. Given these sizes, our data is subject to a bias in which the $w(\theta)$ estimator is biased low compared to the true correlation, given by:

\begin{equation}
\sigma^2=\frac{1}{\Omega^2}\int\int w(\theta)\rm{d}\Omega_1\rm{d}\Omega_2
\end{equation}

\noindent where $\Omega$ is the areal coverage of the data. The bias, known as the Integral Constraint \citep[e.g.][]{1977ApJ...217..385G,1980lssu.book.....P,1993MNRAS.263..360R,1996MNRAS.283L..15B}, results from estimating the mean density from the sample itself. Sampling larger areas reduces this effect, however it remains significant for the size of our survey fields. The `true' $w(\theta)$ is therefore given by: 

\begin{equation}
w(\theta) = \left<w_{meas}(\theta)\right> + \sigma^2
\end{equation}

\noindent where $\left<w_{meas}(\theta)\right>$ is the measured correlation function, averaged across the observed fields, and $w(\theta)$ is the correct correlation function. As in \citet{2002MNRAS.337.1282R}, we evaluate the integral constraint using the numbers of random-random pairs in our fields:

\begin{equation}
\sigma^2=A\frac{\sum N_{RR}(\theta)\theta^{1-\gamma}}{\sum N_{RR}(\theta)}
\end{equation}

For the purposes of this work, we use the commonly used approach of fitting $w(\theta)$ results with a basic power-law of the form:

\begin{equation}
\label{eq:awdel}
w(\theta) = A\theta^{1-\gamma}
\end{equation}

\noindent with the separation angle, $\theta$ in arcminutes. We note that the characterization of galaxy clustering is more and more being treated using the halo-model approach \citep[e.g.][]{2004MNRAS.347..813H,2005ApJ...633..791Z,2008ApJ...682..937B,2008MNRAS.387.1045W}. We restrict the analysis presented here to the power-law fitting, leaving a full halo-modelling analysis to future work. For the real-space clustering we assume the usual power-law form given by:

\begin{equation}
\xi(r) = \left(\frac{r_0}{r}\right)^\gamma
\end{equation}

\noindent where $\xi$ is the real-space two-point correlation function \citep{1980lssu.book.....P} and $r$ is the real-space separation between two points. $r_0$ is then the characteristic separation and $\gamma$ is the power-law slope and is equivalent to the $\gamma$ in Eq.~\ref{eq:awdel}. We fit for the real-space clustering using the analytical transformation from $\xi(r)$ to $w(\theta)$ given by, for example, \citet{phillipps78,1980lssu.book.....P,2007A&A...473..711S}. We note that we do not use Limber's approximation, but use the full form (although the approximation makes little difference given the broadness of the redshift ranges considered when compared to the on-sky coverage).

It is also instructive to estimate the galaxy clustering bias, $b_g$, from the clustering results. This gives the relationship between the clustering of the tracer population, i.e. the selected galaxy samples $\xi_g(r)$, and the underlying dark matter clustering, $\xi_{DM}(r)$.

\begin{equation}
b_g^2 = \frac{\xi_g(r)}{\xi_{DM}(r)}
\end{equation}

We estimate the bias by evaluating the integrated correlation function for both the galaxy clustering and the dark matter halo clustering, which is given by:

\begin{equation}
\overset{\_}{\xi}(r_{max}) = \frac{3}{r_{max}^3}\int_0^{r_{max}}\xi(r)r^2\mbox{d}r
\end{equation}

For the purposes of this study, we use a value of $r_{max}=20h^{-1}$Mpc, which is a large enough scale to apply such that linear theory applies.

\subsection{Clustering of galaxies up to $z\sim2$}

\begin{figure*}
\centering
\includegraphics[width=150.mm]{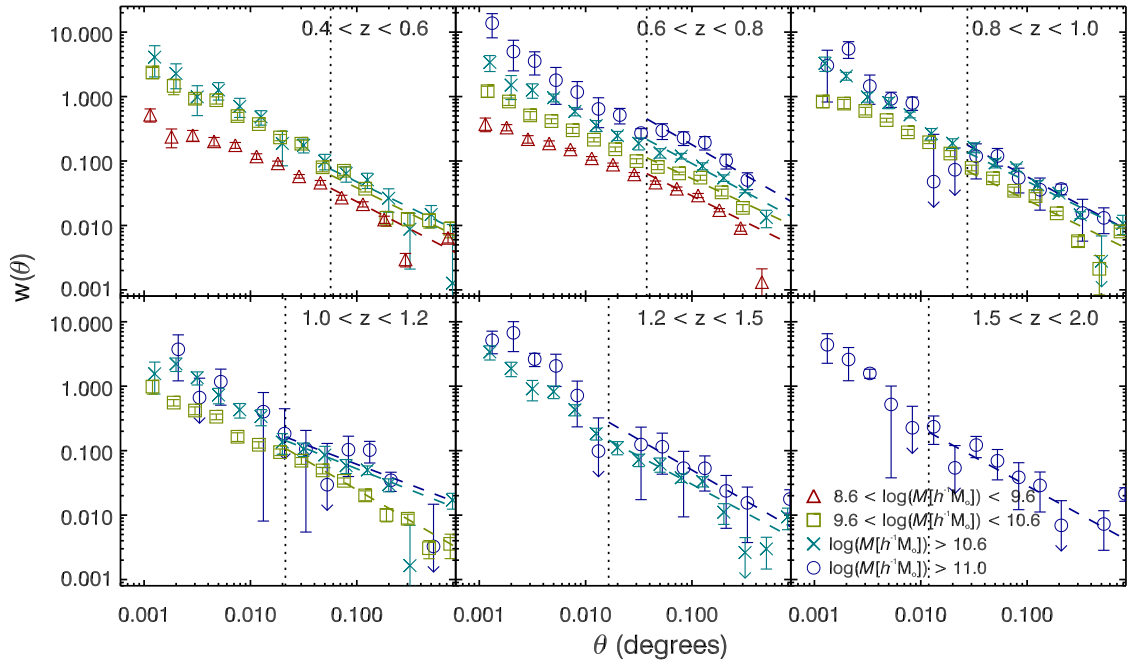}
   \caption{Auto-correlation, $w(\theta)$, of all galaxies up to $z=2$ as a function of mass. The dotted vertical line in each figure gives the $\approx1h^{-1}$Mpc scale for the given redshift range above which the fits are made to the data (i.e. in the 2-halo term regime). The triangles give the $w(\theta)$ measurement for the lowest mass range, $10^{8.6}<M[\hinvm]<10^{9.6}$ range, the squares the $10^{9.6}<M[\hinvm]<10^{10.6}$ range, the $\times$'s the $M[\hinvm]>10^{10.6}$ range and the circles the $M[\hinvm]>10^{11.6}$ range. Errors were estimated using a bootstrap analysis and the dashed lines show the $A_w$-$\gamma$ power law fits.}
   \label{fig:wtheta_all}
\end{figure*}

We first analyse the galaxy population split into bins of mass and redshift. The angular correlation functions are presented in Fig.~\ref{fig:wtheta_all}, with each panel giving the results for a different redshift range as marked. In each panel, the triangles show the $10^{8.6}<M[\hinvm]\leq10^{9.6}$ sample, the squares the $10^{9.6}<M[\hinvm]\leq10^{10.6}$ range, the $\times$ the $M[\hinvm]>10^{10.6}$ range and the circles the $M[\hinvm]>10^{11}$ sample. Note that the correlation function is only plotted where sufficient galaxies in the given bin were present (i.e. $\gtrsim150\degm2$). Each $w(\theta)$ measurement represents the mean of the four fields, whilst the error estimates were calculated using a bootstrap analysis using 100 bootstrap resamplings. In all cases the points include the integral constraint contribution to the clustering.

We perform the power-law fitting to each measurement of the angular correlation function using the form given by Eq.~\ref{eq:awdel}. The fitting is limited to comoving separations of $r>1.2/(1+z)\hmpc$ comoving (given by the vertical dotted lines in each panel in Fig.~\ref{fig:wtheta_all}) in order to minimise the contribution of the non-linear regime (following the example of \citealt{2010MNRAS.406..147F}). The resulting $A_w-\gamma$ fits are given by the dashed lines passing the points in Fig.~\ref{fig:wtheta_all} and the parameters, $A_w$ and $\gamma$ are provided in full in Table~\ref{tab:all_adel}. These power-law profiles successfully fit the data very well within the errors over the scales considered.

From Fig.~\ref{fig:wtheta_all}, a clear link between clustering strength and galaxy stellar mass is evident, with more massive samples showing stronger clustering at all redshifts where multiple samples have been studied.  We note that some of the higher mass clustering results (i.e. at $M[\hinvm]>10^{10.6}$) show some signs of a break indicative of the 2-halo term.

In order to gain a clearer picture of the dependency of the clustering of the galaxy populations on mass and redshift we now estimate the real-space clustering properties of the populations using a single power-law prescription for the real-space correlation function as described in Sec.~\ref{sec:clust_theory}. We therefore determine the clustering length, $r_0$, and the slope, $\gamma$, for each stellar mass and redshift combination (again limiting the fits to just those points at separations of $r>1.2/(1+z)\hmpc$ to minimise the impact of non-linear small scale clustering). The results are given in Table~\ref{tab:r0gam}, with quoted errors estimated from the bootstrap analysis.

\begin{figure}
\centering
\includegraphics[width=90.mm]{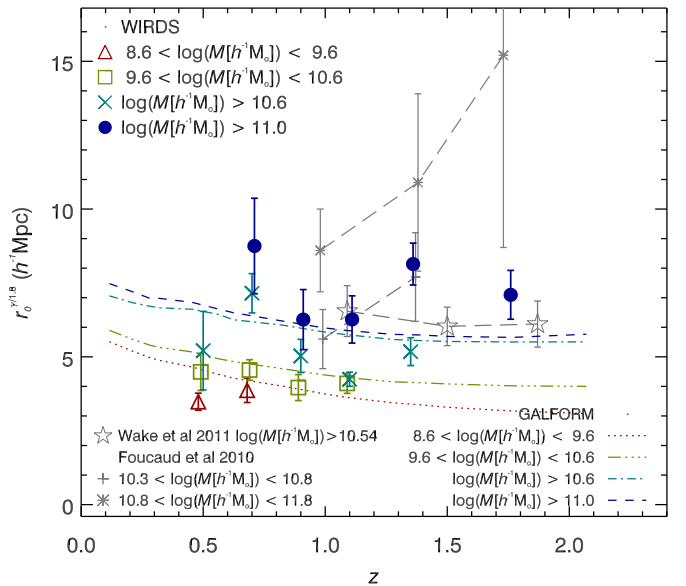}
   \caption{Clustering strength, $r_0^{\gamma/1.8}$ as a function of redshift for all galaxies with errors based on a bootstrap estimate. The populations are split by mass, with triangles showing the $8.6<M[\hinvm]\leq9.6$ samples, squares $9.6<M[\hinvm]\leq10.6$, $\times$'s $M[\hinvm]>10.6$ and circles $M[\hinvm]>11$. The curves give the results of the {\sc galform} model for the observed mass ranges as indicated in the legend. Results from \citet[][grey asterisks and crosses connected by dashed lines]{2010MNRAS.406..147F} and \citet[][grey stars connected by dashed lines]{2011ApJ...728...46W} are also plotted.}
   \label{fig:zvr0gam}
\end{figure}

We show the dependency of $r_0^\gamma$ on redshift and mass in Fig.~\ref{fig:zvr0gam}. The symbols here are the same as in Fig.~\ref{fig:wtheta_all}, with the triangles showing the $10^{8.6}<M[\hinvm]\leq10^{9.6}$ sample, the squares the $10^{9.6}<M[\hinvm]\leq10^{10.6}$ range, the $\times$ the $M[\hinvm]>10^{10.6}$ range and the circles the $M[\hinvm]>10^{11}$ sample. Also plotted are the predictions from the {\sc galform} semi-analytic model described in Sec.~\ref{sec:galform}, with each curve giving the clustering strength as a function of mass, for the mass intervals $10^{8.6}<M[\hinvm]\leq10^{9.6}$ (red dotted curve), $10^{9.6}<M[\hinvm]\leq10^{10.6}$ (green double-dot dash curve), $M[\hinvm]>10^{10.6}$ (turquoise single dot-dash curve) and $M[\hinvm]>10^{11}$ (dashed blue curve). 

The power-law fits affirm the significance of the observed link between stellar mass and the clustering strength across the redshift range. In addition, the $r_0^\gamma$ measurements now illustrate how the clustering evolves with redshift, showing that for a given mass, no significant evolution is seen in $r_0^\gamma$ with redshift up to $z\sim2$. Although there is some tentative indication of the clustering strength increasing from $z\sim1.5$ to $z\sim0.5$ in the $M[\hinvm]>10^{10.6}$ sample, the increase is within the error estimates of the points and no indication of an increase is seen in the lower mass samples. For comparison, we show results from \citet[][grey crosses and asterisks]{2010MNRAS.406..147F} and \citet[][grey stars]{2011ApJ...728...46W}. The \citet{2010MNRAS.406..147F} points show samples with mass ranges of $10^{10.5}<M[\hinvm]<10^{11.0}$ and $10^{11.0}<M[\hinvm]<10^{12.0}$, which also show a significant mass dependence as with our results, but are reported as showing an increase in the clustering strength with redshift. This is at odds with our results, however we note that the \citet{2010MNRAS.406..147F} measurements of equivalent mass bins give consistent clustering strengths within the errors and so the increase does not appear to be a significant one. Indeed, the \citet{2011ApJ...728...46W} measurements are more tightly constrained and agree well with our results, both in terms of the magnitude of the measured clustering and the overall trend with redshift of no significant redshift evolution in $r_0^{\gamma/1.8}$. The WIRDS data is $\approx$ 1 magnitude deeper than the \citet{2010MNRAS.406..147F} data and covers approximately twice the area (2.4 deg$^2$ compared to 1.16 deg$^2$). The additional depth of the WIRDS data is particularly significant in the highest redshift bins, where we are likely probing a much greater proportion of the lower mass end of the $M>10^{11}\rm{M_\odot}$ range than \citet{2010MNRAS.406..147F}.

Comparing the observations with the {\sc galform} predictions, we find that the predicted mass dependence agrees with our observations. However, the {\sc galform} model marginally over-predicts the clustering when compared with the observations, except for the highest mass bins. The model predicts an upturn in the clustering towards lower redshift (i.e. $z\lesssim1$). This is not seen with any significance in the observations, however it can not be ruled out by them either.

\begin{figure}
\centering
\includegraphics[width=80.mm]{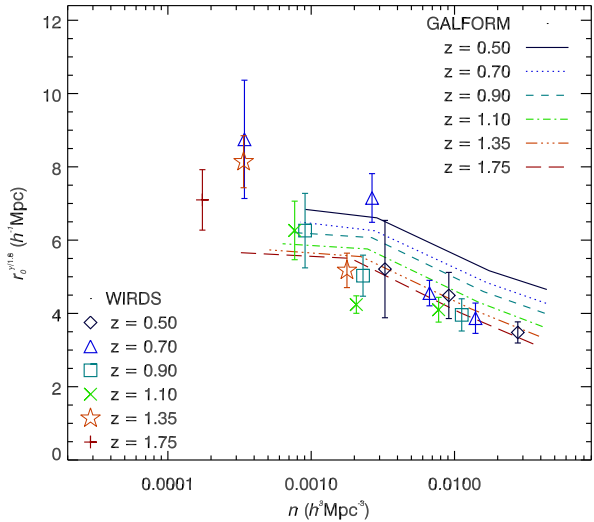}
   \caption{Clustering strength, $r_0^{\gamma/1.8}$ as a function of number density. The points show the results from the WIRDS analysis separated by redshift. The lines show the equivalent trends, again separated by redshift, from the {\sc galform} model. The trend of increasing clustering strength with decreasing space density is significant in the results.}
   \label{fig:nvr0gamall}
\end{figure}

Fig.~\ref{fig:nvr0gamall} shows the relation between the clustering strength and the space densities of these populations. The points show the results for the WIRDS galaxy samples, with the black diamonds showing measurements at $z=0.5$, blue triangles showing results for $z=0.7$, cyan squares the results for $z=0.9$, the green times-symbols $z=1.1$, the red stars $z=1.35$ and red crosses $z=1.75$. Number densities were calculated based on the galaxy stellar mass function calculated in the same redshift bins by \citet{2012A&A...545A..23B}. The dependence of clustering on galaxy space density is clear and significant, with rarer galaxies being more highly clustered. This is strongly connected to the stellar masses of the galaxy samples and the results presented in Fig.~\ref{fig:zvr0gam} where galaxy populations with higher masses are more strongly clustered. The trends predicted with the semi-analytical model are comparable to the observations. However the model over-predicts the clustering measurements of the galaxy populations. We also note that the model predicts a redshift evolution, with the trend moving to higher clustering strengths with lower redshift and that this trend with redshift is not seen in the observational results. This is particularly the case at high number densities ($n\gtrsim0.004\hcmpcc$), where a tight line of data-points is seen compared to the evolution of the model curves. This appears to be at least in part driven by the over-prediction in the numbers of low mass galaxies seen in the galaxy stellar-mass functions presented in \citet{2012A&A...545A..23B} for this same data and model combination.

\subsection{Clustering of galaxies by type}

We now perform the same analysis, but with the sample split into passive and star-forming galaxy populations. The sample is split based on derived dust-corrected rest-frame $NUV-R$ colours. The selection reliably differentiates galaxies based on star-formation rates in the sample to $z\sim2$ and is described in detail by \citet{2010ApJ...709..644I} and \citet{2012A&A...545A..23B}. We apply an identical cut to the {\sc galform} model galaxies, producing measurements of the model galaxy clustering in an identical manner to that used for the observed galaxy populations.

\begin{figure}
\centering
\includegraphics[width=80.mm]{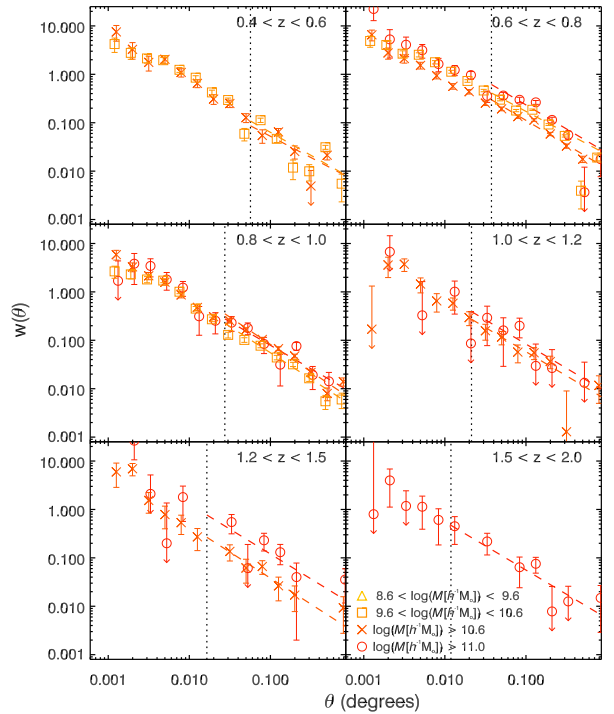}
   \caption{As in Fig.~\ref{fig:wtheta_all}, but for only the passive galaxy population selected by rest-frame dereddenned $M_{NUV}-M_r$ colour.}
   \label{fig:wtheta_red}
\end{figure}

\begin{figure}
\centering
\includegraphics[width=80.mm]{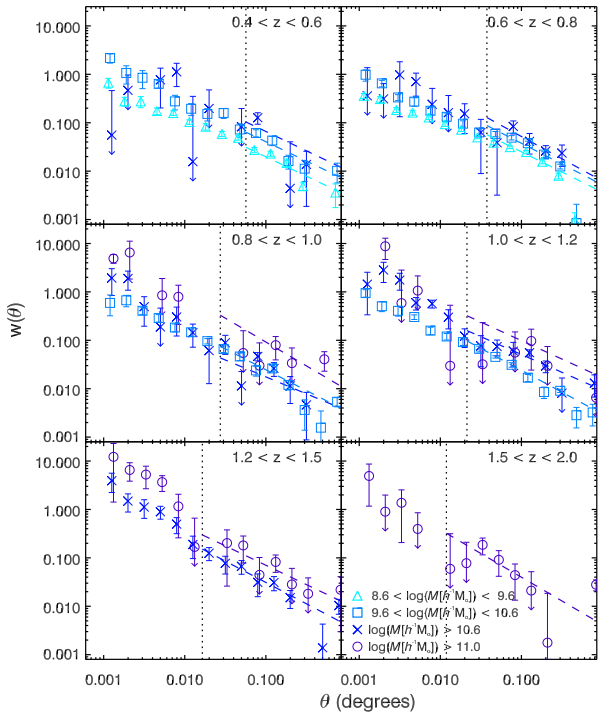}
   \caption{As in Fig.~\ref{fig:wtheta_all}, but for only the star-forming galaxy population selected by rest-frame dereddenned $M_{NUV}-M_r$ colour.}
   \label{fig:wtheta_blue}
\end{figure}

The clustering results for passive and star-forming galaxies are shown in Figs.~\ref{fig:wtheta_red} and \ref{fig:wtheta_blue} respectively. We again fit the $w(\theta)$ measurements with $A_w\theta^{1-\gamma}$ power-laws, the results of which are plotted as dashed lines in Figs.~\ref{fig:wtheta_red} and \ref{fig:wtheta_blue} and recorded in Table~\ref{tab:all_adel}. Again the galaxy populations in each redshift range are split by mass, with triangles denoting the $10^{8.6}<M[\hinvm]\leq10^{9.6}$ bin, squares the $10^{9.6}<M[\hinvm]\leq10^{10.6}$ bin, $\times$ the $M[\hinvm]>10^{10.6}$ and the circles give the results for the $M[\hinvm]>10^{11}$ bin. The same trend of increasing clustering strength with increasing mass is seen for the star-forming population as in the full population. However, the results for the passive galaxies seem less clear, with the auto-correlations of the $10^{9.6}<M[\hinvm]\leq10^{10.6}$ and $M[\hinvm]>10^{10.6}$ being largely consistent with each other in the redshift ranges where both are measured.

\begin{figure}
\centering
\includegraphics[width=90.mm]{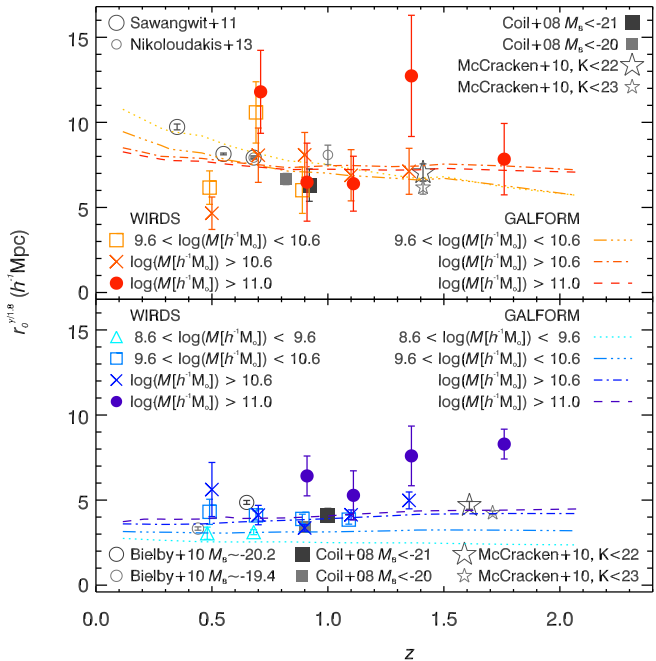}
   \caption{Clustering strength, $r_0^{\gamma/1.8}$ as a function of redshift for passive galaxies (top panel) and star-forming galaxies (bottom panel). In each case, the populations are split by mass, with triangles showing the $8.6<M[\hinvm]\leq9.6$ range, squares $9.6<M[\hinvm]\leq10.6$, $\times$'s $M[\hinvm]>10.6$ and circles $M[\hinvm]>11$. The curves give the predictions of the {\sc galform} model for the different mass bins as indicated in the legend.}
   \label{fig:zvr0gam_bytype}
\end{figure}

Again we calculate real-space correlation function power-law fits to the angular auto-correlation functions. The resulting $r_0$ and $\gamma$ parameters are given in Table~\ref{tab:r0gam} and plotted in Fig.~\ref{fig:zvr0gam_bytype}, with passive galaxies plotted in the top panel and star-forming in the lower panel. It is clear from both the plots and the parameter values that, for a given mass bin and redshift, the passive galaxy samples are significantly more clustered than the star-forming galaxy populations. This dependency has been well probed at $z\lesssim1$ \citep[e.g.][]{1976ApJ...208...13D,1981MNRAS.194...49P,1995ApJ...442..457L,2002MNRAS.332..827N}, but results remain ambiguous at $z\gtrsim1$ \citep[e.g.][]{2006A&A...452..387M,2008ApJ...672..153C,2010MNRAS.407.1212H}. Our observations are consistent with passive galaxies being more likely to exist in dense environments such as clusters and groups across the entire redshift range probed here (i.e. to $z=2$), with the star-forming populations having clustering strengths of $r_0^\gamma\lesssim5\hmpc$ (consistent with comparable measures made at $z\lesssim1$, e.g. \citealt{2002MNRAS.332..827N,2009MNRAS.395..240B,2010MNRAS.403.1261B}). This suggests that the process that has produced this difference in the clustering of passive and star-forming galaxies has already been at work much before $z\sim2$.

We note however, that the most massive star-forming galaxies are seen to have clustering strengths of $r_0^{\gamma/1.8}\sim7-8\hmpc$, approaching the clustering strengths of the lowest mass passive galaxy samples. This is a similar result as that seen by \citet{2010MNRAS.407.1212H}, whereby they report clustering amplitudes for star-forming galaxies (with absolute $K$-band magnitudes of $M_K\sim-25$) at $z\gtrsim1.5$ comparable to those of passive galaxy samples. \citet{2010MNRAS.407.1212H} suggest that this increase is indicative of star-forming galaxies being found in more highly clustered environments at high-redshift. Indeed, the authors conclude that the clustering strengths of star-forming galaxies decline towards $z=0$ for a given $K$-band luminosity (where this is a proxy for stellar mass). Our results could suggest a similar trend towards higher clustering at higher redshift for the star-forming galaxies, but when taking into account the error estimates, the clustering measurements are consistent at the $\sim1\sigma$ level for all the redshift bins. We note also that the massive star-forming galaxies being as highly clustered as the passive population is consistent with the results of \citet{2013arXiv1308.2974T} at $z\lesssim1$ in which it is seen that star-forming galaxies above a mass limit of $M_\star~10^{10.8}M_\odot$ have large scale clustering amplitudes comparable with those of the passive galaxy population. Significantly, our results provide new evidence for the continuation of this trend to $z\gtrsim1$.

In Fig.~\ref{fig:zvr0gam_bytype}, we plot results from a number of other studies of the clustering of star-forming galaxies across the redshift range we consider. The open circles show the clustering of photometrically selected star-forming galaxies at low redshift \citep{2010MNRAS.403.1261B}, the filled squares show comparable results from the DEEP2 survey presented by \citet{2008ApJ...672..153C} at $z\sim1$ and the open stars show results for star-forming $BzK$ selected galaxies in the COSMOS field presented by \citet{2010ApJ...708..202M}. These samples add to the picture of little overall change in the clustering strength of star-forming galaxies over time since $z\sim2$. Indeed, results for spectroscopic samples of LBGs at $z\sim3$, which are broadly representative of the luminous star-forming population, give clustering lengths of $r_0^{\gamma/1.8}\sim4\hmpc$ \citep[e.g.][]{adelberger03,2005ApJ...619..697A,2011MNRAS.414....2B,2013MNRAS.430..425B}. These are estimated to have mean stellar masses of $M_\star\sim10^{10.3}~h^{-1}{\rm M_\odot}$ \citep{shapley03} at which mass we measure clustering lengths of $3.8\hmpc<r_0<4.5\hmpc$, adding to our observation that there is little evolution in the clustering as measured via $r_0^{\gamma/1.8}$ as a function of redshift to $z\sim2-3$.

Returning to the passive galaxies, all but two of the observational points agree at the $1\sigma$ level with the {\sc galform} predictions. Both observations and the model predictions suggest little change in clustering strength with redshift. This lack of evolution with redshift is consistent with previous results at lower redshift. In the redshift range $0.4\lesssim z\lesssim1.0$, LRG observations (of samples of approximately uniform absolute magnitude) are consistent with no evolution of $r_0^{\gamma/1.8}$ as a function of redshift \citep[][shown in Fig.~\ref{fig:zvr0gam_bytype}]{2011MNRAS.416.3033S,2012arXiv1204.3609N}. The WIRDS data are consistent with the LRG results, but extend this conclusion to lower mass samples than the LRGs and to higher redshift.

Returning to the stellar mass dependence of the clustering for passive galaxies, as stated we find that the data show no significant correlation here. In fact, the lower mass samples are found to have higher clustering sxtrengths than the higher mass samples at the lower redshifts ($z\lesssim0.8$) in our study. This is similar to the results reported by \citet{2013MNRAS.431.3045H} for clustering with stellar mass and \citet{2009ApJ...691.1879W} for clustering as a function of luminosity for passive galaxies. In fact we note that over the absolute luminosity range we probe with our samples (i.e. $-16\gtrsim M_g\gtrsim-22$), this is consistent with previous work at low redshift such as \citet{2002MNRAS.332..827N} in which a small dependence of clustering on absolute magnitude is seen at $M_{bj}\gtrsim-21$. Further to this significant correlations between clustering strength and luminosity are reported for higher luminosity systems \citep[e.g.][]{2011MNRAS.416.3033S}. Combining our own results with \citet{2013MNRAS.431.3045H} and the lower redshift results, we conclude that similarly to the luminosity dependence shown by \citet{2002MNRAS.332..827N}, the clustering of passive galaxies has little dependence on stellar mass below masses of $M_\star\lesssim10^{11}~h^{-1}{\rm M}_\odot$, but becomes more dependent on stellar mass above this limit. The {\sc galform} model also predicts only a very small dependency of $r_0^{\gamma/1.8}$ on $M_*$ for passively evolving galaxies. This is found to be the result of the halo mass being, on average, a constant with galaxy $M_*$ in the model.

The result of an observed red-sequence with high clustering levels to $z\approx2$ is complimentary and consistent with observations of the red-sequence to be in place in galaxy clusters to such redshfits \citep[e.g.][]{2010ApJ...716L.152T,2011A&A...526A.133G} and the lack of evolution in these red-sequence galaxies and brightest cluster galaxies \citep[e.g.][]{2010ApJ...715L...6O,2011MNRAS.414..445S,2012A&A...545A..23B}.

The results for the star-forming populations are shown in the lower panel of Fig.~\ref{fig:zvr0gam_bytype}. Here again we see little sign of evolution in the clustering strength over the redshift range probed. The highest redshift measurement shows some sign of an upturn and is indeed as comparable to the equivalent measurement for the passive galaxy population at this redshift. However, the increase compared to the lower-redshift measurements of $r_0^{\gamma/1.8}$ is only at the 1$\sigma$ level. As discussed, the most massive star-forming galaxies at $z>1.2$ do lie in strongly clustered regions and have clustering strengths comparable to low mass passive galaxies. We note that this epoch coincides with recent claims of significant star-formation rates in high redshift clusters. For example, \citet{2010ApJ...718..133H} report observations of 24 $\mu m$ sources within $<$ 250 kpc of the centre of the high redshift cluster XMMXCS J2215.9-1738 ($z=1.46$), which they report suggests that a large amount of star formation may be taking place in the cluster core, in contrast to clusters at lower redshifts. Similarly, \citet{2010ApJ...719L.126T} measure an increase in the fraction 24 $\mu m$ luminous star-forming galaxies towards the centre of the $z=1.62$ cluster CIG J0218.3-0510, again in contradiction to results at lower redshift and signifying a shift in the location of star-formation to the high density regions.

We again see good agreement between the observations and the model, with both showing relatively constant clustering strengths as a function of redshift for each mass range. We also observe a stronger mass dependence at all redshifts for the star-forming populations than for the passive galaxies. This is again the case for both the observational results and the {\sc galform} model predictions. For clarity we note that the lowest dashed line corresponds to the same mass range as the triangular points, the second line to the square points, the third to the $\times$ and the highest to the mass range of the filled circles. The star-forming population therefore appears to be the dominant contributor to the increase in clustering with galaxy stellar mass content, whilst the passive galaxies show a smaller, less clear, change in clustering with stellar mass within the mass ranges we are probing.

\begin{figure}
\centering
\includegraphics[width=90.mm]{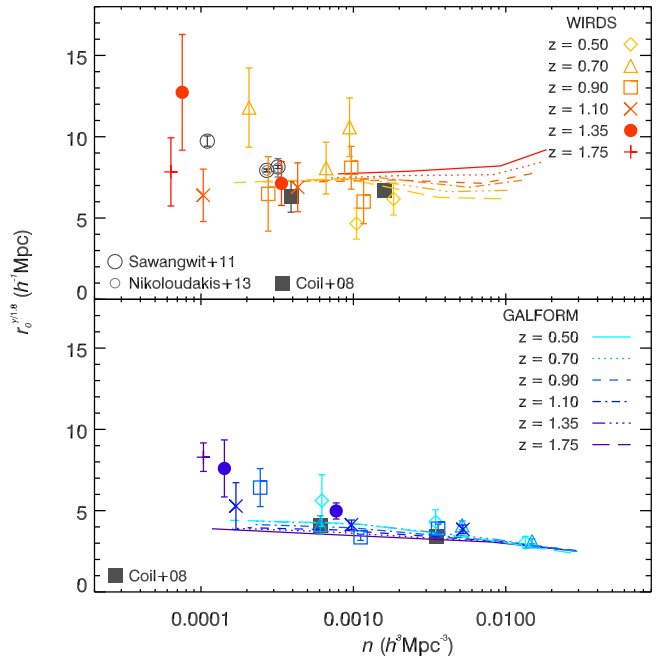}
   \caption{Clustering strength, $r_0^{\gamma/1.8}$ as a function of number density, split by galaxy type, with passive galaxies shown in the top panel and star-forming galaxies in the lower panel. In both panels, the coloured symbols show the results for the WIRDS data, whilst the lines show the predictions from {\sc galform}. The grey symbols show results from the literature, with the $z=0.3,0.5,0.7$ LRG results from \citet[][large open circles]{2011MNRAS.416.3033S}, the $z=0.7,1.0$ LRG results from \citet[][small open circles]{2012arXiv1204.3609N} and the passive (plotted top) and star-forming (plotted in the bottom panel) $z=1$ results of \citet[][filled squares in both panels]{2008ApJ...672..153C}.}
   \label{fig:nvr0gambytype}
\end{figure}

In Fig.~\ref{fig:nvr0gambytype}, we show the clustering results for the passive (top) and star-forming (bottom) galaxies as a function of sample number density. In each case the points are split by redshift, with the diamonds corresponding $z=0.5$, the triangles to $z=0.7$, the squares to $z=0.9$, the times symbols to $z=1.1$, the stars to $z=1.35$ and the crosses to $z=1.75$. Interestingly, the {\sc galform} predictions suggest little dependence of clustering strength on galaxy number density for the passive galaxies. This is closely related to the small galaxy stellar mass dependence seen in the previous figure. Using the WIRDS data alone it is difficult to constrain this prediction, however we can add in results from previous work to aid the analysis. As such we include the previous points for LRGs, which represents a strongly clustered low number-density population, and the passive galaxies analysed by \citet{2008ApJ...672..153C}. These are consistent with the WIRDS points, with the \citet{2008ApJ...672..153C} points in particular corroborating the lack of evolution seen in the {\sc galform} predictions, whilst the \citet{2011MNRAS.416.3033S} results suggest some dependence of $r_0^{\gamma/1.8}$ at number densities of $\lesssim3\times10^{-4}~\hcmpcc$.

A more clear variation of $r_0^{\gamma/1.8}$ is seen with number density in the lower panel, where the results are shown for star-forming galaxies. In comparison to the passive population, far less scatter is seen in the results, due to the larger numbers of star-forming galaxies and the resulting improvement in statistical errors. Again we show points from \citet{2008ApJ...672..153C} and find good agreement between their results for star-forming galaxies and our own. The {\sc galform} model predicts a relation between $r_0^{\gamma/1.8}$ and number density that is consistent with the WIRDS results down to number densities of $\sim2\times10^{-4}~\hcmpcc$, where the observational data suggest a potential upturn in the trend.

\subsection{Dark Matter Halo Mass}

From the clustering results, we may estimate the mean masses of dark matter halos within which the galaxy samples reside. We now do this using the formalism developed by \citet{mowhite96}. This provides a relationship between the bias of galaxy clustering to the mean halo mass, based on a model of spherical collapse and tested with N-body simulations. An extension to the method was made based on ellipsoidal collapse by \citet{2001MNRAS.323....1S}, which relates the halo-bias to the mean halo mass via:

\begin{eqnarray}
b_h(M_{\rm DM},z) & = & 1 + \frac{1}{\sqrt{a}\delta_c}\left[a\nu^2\sqrt{a}+b\sqrt{a}(a\nu^2)^{1-c} \right. \nonumber \\
           & & \left. - \frac{(a\nu^2)^c}{(a\nu^2)^c+b(1-c)(1-c/2)}\right]
\label{eq:biasmdm}
\end{eqnarray}

\noindent where $a$, $b$ and $c$ are constants for which we take the values given by \citet{2005ApJ...631...41T}: $a=0.707$, $b=0.35$ and $c=0.8$. $\delta_c$ is the critical overdensity required for collapse and is given by $\delta_c=0.15(12\pi)^{2/3}\Omega_m(z)^{0.005}\approx1.686$ \citep{1997ApJ...490..493N}. The variable $\nu$ is defined as $\delta_c/\sigma(M_{\rm DM},z)$, where $\sigma(M_{DMH,z})$ is the rms fluctuation of the density field and can be separated into dark matter halo mass and redshift dependancies via $\sigma(M_{\rm DM},z)=\sigma(M_{\rm DM})D(z)$. Here $D(z)$ is the linear growth rate and the mass dependence of the rms fluctuation is given by:

\begin{equation}
\sigma(M_{\rm DM})^2 = \frac{1}{2\pi^2}\int_0^\infty k^2P(k)w(kr)^2\mbox{d}k
\label{eq:sigmdm}
\end{equation}

Here, $P(k)$ is the matter power-spectrum, which we calculate using \texttt{CAMB} \citep{lewis00,2011PhRvD..84d3516C}, which is based on \texttt{CMBFAST} \citep{1996ApJ...469..437S,2000ApJS..129..431Z}. $w(kr)$ is the window function for a spherical top-hat function given by:

\begin{equation}
w(kr) = 3\frac{\sin(kr)-kr\cos(kr)}{(kr)^3}
\label{eq:tophat}
\end{equation}

\noindent where $r$ is the top-hat radius and is related to the mass, $M_{\rm DM}$ by:

\begin{equation}
r = \left(\frac{3M_{\rm DM}}{4\pi\rho_0}\right)^\frac{1}{3}
\label{eq:tophatrad}
\end{equation} 

$\rho_0$ is the present day mean density of the universe and is given by $\rho_0=\Omega_m^0\rho_{crit}^0=2.78\times10^{11}\Omega_m^0h^2M_\odot\mbox{Mpc}^{-3}$.

Combining equations \ref{eq:biasmdm}, \ref{eq:sigmdm}, \ref{eq:tophat} and \ref{eq:tophatrad} allows us to estimate the dark matter halo mass from the clustering bias, which we match to the calculated bias for each of the galaxy samples.

\begin{figure}
\centering
\includegraphics[width=90.mm]{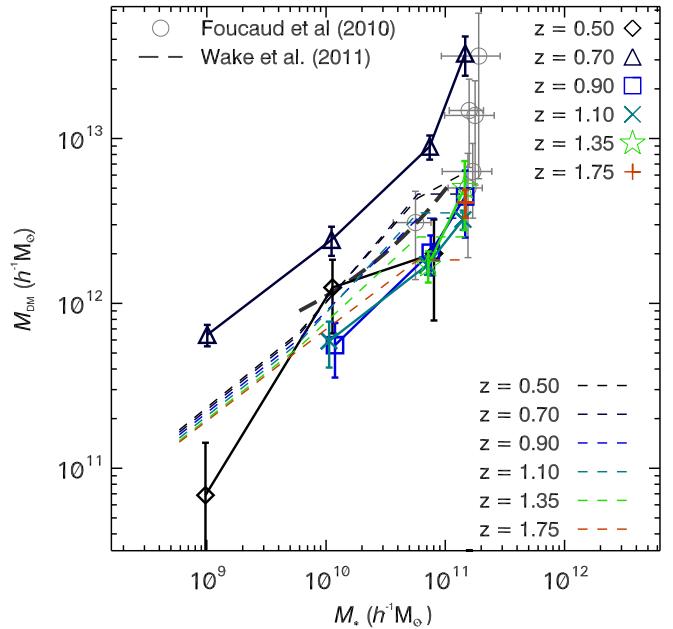}
   \caption{Host dark matter halo mass, $M_{\rm DM}$, as a function of galaxy stellar mass, $M_{*}$ for the full galaxy sample split by mass and redshift. The triangles, squares, $\times$ and stars show the results from the WIRDS data for redshift ranges centred on $z=0.45$, $z=0.80$, $z=1.20$ and $z=1.70$ respectively. Predictions from the {\sc galform} model are given for the same central redshifts. We also show the fit to $M_{\rm DM}$ versus $M_min$ given by \citet{2011ApJ...728...46W} for galaxies at $1<z<2$ in the NEWFIRM survey.}
   \label{fig:all_hmass}
\end{figure}

We show the results of the bias matching in Fig.~\ref{fig:all_hmass}, where we plot the minimum dark matter halo mass, $M_{\rm DM}$, versus the mean galaxy sample stellar mass, $M_\ast$. This is based on the full galaxy sample clustering as presented in Fig.~\ref{fig:zvr0gam}. The WIRDS data is split by redshift with the black diamonds showing the $z=0.5$ points, the dark blue triangles showing the $z=0.7$ points, the blue squares showing the $z=0.9$ points, the cyan times symbols the $z=1.1$ points, the green stars the $z=1.35$ points and the red crosses the $z=1.75$ points. At each redshift, a clear dependence of $M_{\rm DM}$ on stellar mass is seen. In addition, there is some sign of a redshift evolution, with $M_{\rm DM}$ appearing to move lower for a given stellar mass with increasing redshift.

Comparing to equivalent datasets in the literature, the grey circles show the points of \citet{2010MNRAS.406..147F} and the long-dashed curve shows a fit from \citet{2011ApJ...728...46W}. The \citet{2010MNRAS.406..147F} points are based on equivalent redshift and mass bins to our highest redshift bins. As in Fig.~\ref{fig:zvr0gam}, the WIRDS and \citet{2010MNRAS.406..147F} results agree well except where the latter find large clustering results: in this case halo masses of $M_{\rm DM}\gtrsim10^{13}~h^{-1}{\rm Mpc}$. Our results at these masses and redshifts (green stars \& red crosses) suggest much lower halo masses $M_{\rm DM}\gtrsim4-8\times10^{12}~h^{-1}{\rm Mpc}$. However, given the large error bars on the \citet{2010MNRAS.406..147F} points, their results are consistent with those of WIRDS as well as the relation derived in \citet{2011ApJ...728...46W}.

Now taking the {\sc galform} predictions, we find that these agree with the WIRDS results to within $\approx1\sigma$ of the data points across the different redshift ranges, excepting the $z=0.7$ WIRDS results. As discussed previously, the $z=0.7$ data-points seem to be pushed to higher clustering due to the number of large clusters in this redshift range in the COSMOS field. Interestingly, we see that the {\sc galform} results predict a redshift evolution in the halo masses for a given stellar mass cut. The range covered by the WIRDS results and the associated errors limit our ability to confirm whether this is a genuine evolution. We also note that in the model, the relationship between $M_{DM}$ and $M_\star$ is primarily driven by the star-forming population, whilst the halo masses of passive galaxies show little dependence on the stellar mass of the galaxies.

\begin{figure}
\centering
\includegraphics[width=90.mm]{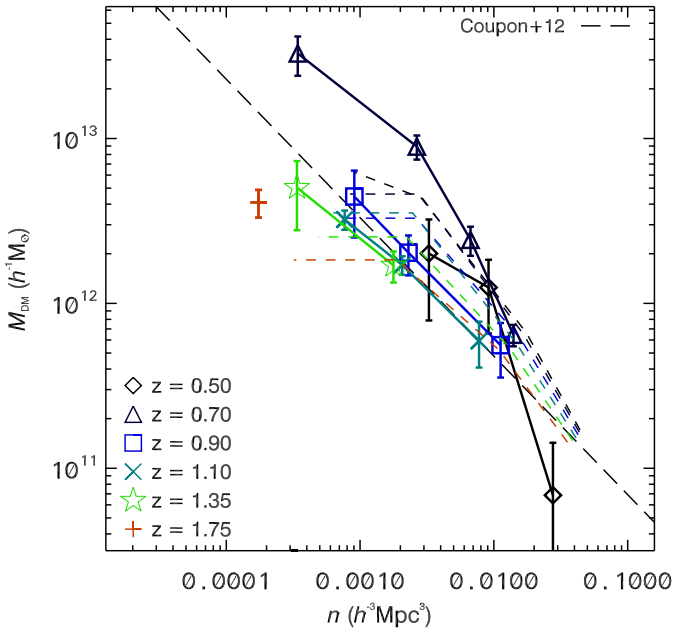}
   \caption{The mean halo mass versus galaxy number density for our range of galaxy stellar mass and redshift ranges. The points show our results withe different point types giving different redshift ranges. The long-dashed line shows the best fit determined by \citet{2012A&A...542A...5C} for $z<1$ galaxies in the CFHTLS. The short-dashed curves show the {\sc galform} predictions with the same colour coding as given in Fig.~\ref{fig:all_hmass}.}
   \label{fig:hmassnumdens}
\end{figure}

Taking our estimates for halo masses from the clustering results, we now plot these against number density, $n$, in Fig.~\ref{fig:hmassnumdens}. We use the same symbols for different redshift bins as in Fig.~\ref{fig:all_hmass}. Also plotted is the $n-M_{\rm DM}$relation determined for low-redshift ($z\lesssim1$) galaxies by \citet{2012A&A...542A...5C} from the CFHTLS Wide field data. Again, the {\sc galform} results are shown by the short dashed coloured lines, with the colours coordinated with the WIRDS data points (i.e. from black for low redshift to red for high redshift).

One point to note from the WIRDS data is that plotted in this way, the $z=0.7$ points are consistent with the $z=0.5$ and $z=0.9$ results. However, above $z\sim1$, we now see a tentative trend for the WIRDS results to shift to lower halo masses for a given number density. This effect is also present in the {\sc galform} predictions, with the dashed lines moving down and to the left (note that the {\sc galform}  results are based on the same stellar mass bins as the data points) with increasing redshift.

In summary, from the halo mass analysis, we confirm the relationship between galaxy stellar mass and halo mass across the redshift range of $0.4\lesssim z\lesssim2.0$ in agreement with previous work, whilst finding tentative evidence for an evolution in the relationship between number density and halo mass.

\section{Conclusions}
\label{sec:conclusions}

Based on the deep 8-band CFHTLS$+$WIRDS photometry, we have conducted an analysis of clustering up to $z=2$, evaluating galaxy spatial correlations as a function of mass and type. Complimentary to this, we have used the {\sc galform} semi-analytical model of galaxy formation and evolution to make clustering predictions using identical selection constraints based on the same galaxy properties. From the WIRDS data, we find a constant clustering strength within the error estimates as a function of redshift over the redshift range $0.3<z<2$ for the full galaxy sample. At the same time we detect a mass dependence for clustering over the whole range such that galaxies with greater stellar masses are more strongly clustered, extending previous results to higher redshifts. The {\sc galform} model predictions are consistent with the results from the data.

Comparing the clustering between star-forming and passive galaxies, we again find that both remain constant in clustering strength within the errors as a function of redshift (in the range $0.3<z<2$) for a given mass limit. Thus the `passivity' dependence of clustering, such that passive galaxies are more strongly clustered than star-forming ones, is confirmed up to $z\approx2$. This is only broken at high stellar mass ($M_\star\gtrsim10^{11}\hinvm$) at which point we find star-forming galaxies have comparable clustering (at the large scales probed) to passive galaxies. The {\sc galform} model predictions for the same mass and redshift constraints reproduce the observations well within the observational errors, although do not predict the high clustering found in the high mass star-forming galaxy population.

We find little dependence on mass for the clustering of the passive population across the stellar mass range covered here. This is consistent with the dependence with $b_j$-band magnitude of galaxies brighter than $M_{bj}<-21$ \citep{2002MNRAS.332..827N} (a range within which the majority of our galaxy mass samples are within). Taking the star-forming population, a stronger dependence of clustering strength on stellar mass is observed than for the passive galaxies, a trend that is seen in both the WIRDS data and the {\sc galform} model predictions.

Finally we have investigated the relation between galaxy stellar mass and mean dark matter halo mass for the samples described above. Based on the \citet{mowhite96} formalism, we have estimated mean dark matter halo masses from the clustering measurements for the full galaxy sample. We see a significant trend of mean halo mass increasing with galaxy stellar mass across a range of redshifts. Additionally, we find the tentative result that given a constant stellar mass the halo mass increases marginally with decreasing redshift.

The above results have built on the current picture of galaxy evolution via clustering analyses, adding to the previous works of \citet{2007MNRAS.376L..20F} and \citet{2008MNRAS.387.1045W} with deeper limits in magnitude and mass over an overall larger area. The key results that the passive and star-forming populations remain relatively constant in terms of clustering strength as a function of redshift up to $z\approx2$ (given a constant mass selection) fits in well with complimentary observations showing little evolution in the stellar mass function over a similar range \citep{2010ApJ...709..644I}. It will now be interesting to build on these observations by pushing further down the stellar mass function at $z\gtrsim1$ with deeper and wider NIR data from the upcoming surveys such as UltraVISTA, to constrain the clustering and hence galaxy evolution for galaxies across a broader range of stellar masses.

\begin{acknowledgements}
RMB acknowledges the funding of the French Agence Nationale de la Recherche (ANR) and the UK Science and Technology Facilities Council.  H. J. McCracken acknowledges support from the ``Programme national cosmologie et galaxies". This work is based in part on data products produced at TERAPIX at the Institut d'Astrophysique de Paris and the Canadian Astronomy Data Centre (CADC) as part of the Canada-France-Hawaii Telescope (CFHT) Legacy Survey, a collaborative project of the Canadian National Research Council (NRC) and Centre National de la Recherche Scientifique (CNRS). Much of the work is based on observations at the CFHT, which is funded by the NRC, the CNRS, and the University of Hawaii. This research has made use of the VizieR catalogue access tool provided by the CDS, Strasbourg, France. This research was supported by ANR grant ``ANR-07-BLAN-0228''. ED also acknowledges support from ``ANR-08-JCJC-0008''.  JPK acknowledges support from the CNRS. {\sc galform} was run on the ICC Cosmology Machine, which is part of the DiRAC Facility jointly funded by STFC, the Large Facilities Capital Fund of BIS, and Durham University.
\end{acknowledgements}

\bibliographystyle{aa}
\bibliography{$HOME/Documents/lib/rmb}

\appendix
\section{Results of the $w(\theta)$ fitting}

We give the parameters obtained for the fitting of the angular correlation function results for all combinations of galaxy mass, redshift and type in table~\ref{tab:all_adel} ($A_w$-$\gamma$ fits) and table~\ref{tab:r0gam} ($r_0$-$\gamma$ fits). 

\begin{table*}
\caption{Fit parameters, $A_w$ (in units of arcmin$^{1-\gamma}$) and $\gamma$, for the correlation functions of all galaxies as a function of mass and redshift.}
\label{tab:all_adel}
\centering          
\begin{tabular}{lcccccccccc}     
\hline\hline       
$z$ &  \multicolumn{2}{c}{$10^{8.6}<M[\hinvm]\leq10^{9.6}$} & \multicolumn{2}{c}{$10^{9.6}<M[\hinvm]\leq10^{10.6}$}& \multicolumn{2}{c}{$M[\hinvm]>10^{10.6}$} & \multicolumn{2}{c}{$M[\hinvm]>10^{11}$} \\
         & $A_w$ & $\gamma$& $A_w$ & $\gamma$& $A_w$ & $\gamma$& $A_w$ & $\gamma$\\
\hline
\multicolumn{9}{l}{Passive$+$Star-forming} \\
0.50
& $0.12^{+0.02}_{-0.02}$ & $1.90^{+0.08}_{-0.08}$
& $0.17^{+0.04}_{-0.04}$ & $1.80^{+0.10}_{-0.10}$
& $0.31^{+0.16}_{-0.16}$ & $2.00^{+0.26}_{-0.26}$
& --- & --- 
\\[0.08cm]
0.70
& $0.12^{+0.02}_{-0.02}$ & $1.79^{+0.03}_{-0.03}$
& $0.17^{+0.02}_{-0.02}$ & $1.68^{+0.06}_{-0.06}$
& $0.40^{+0.07}_{-0.07}$ & $1.84^{+0.06}_{-0.06}$
& $0.48^{+0.15}_{-0.15}$ & $1.66^{+0.09}_{-0.09}$
\\[0.08cm]
0.90
& --- & --- 
& $0.11^{+0.02}_{-0.02}$ & $1.80^{+0.08}_{-0.08}$
& $0.21^{+0.04}_{-0.04}$ & $1.80^{+0.09}_{-0.09}$
& $0.24^{+0.07}_{-0.07}$ & $1.83^{+0.20}_{-0.20}$
\\[0.08cm]
1.10
& --- & --- 
& $0.12^{+0.02}_{-0.02}$ & $1.90^{+0.04}_{-0.04}$
& $0.16^{+0.01}_{-0.01}$ & $1.63^{+0.13}_{-0.13}$
& $0.28^{+0.06}_{-0.06}$ & $1.72^{+0.00}_{-0.00}$
\\[0.08cm]
1.35
& --- & --- 
& --- & --- 
& $0.14^{+0.02}_{-0.02}$ & $1.86^{+0.09}_{-0.09}$
& $0.44^{+0.07}_{-0.07}$ & $1.82^{+0.00}_{-0.00}$
\\[0.08cm]
1.75
& --- & --- 
& --- & --- 
& --- & --- 
& $0.16^{+0.04}_{-0.04}$ & $1.94^{+0.21}_{-0.21}$
\\[0.08cm]
\hline
Passive  & & & & & & & & \\
0.50
& --- & --- 
& $0.34^{+0.11}_{-0.11}$ & $2.00^{+0.17}_{-0.17}$
& $0.34^{+0.14}_{-0.14}$ & $1.99^{+0.26}_{-0.26}$
& --- & --- 
\\[0.08cm]
0.70
& --- & --- 
& $0.83^{+0.27}_{-0.27}$ & $1.88^{+0.07}_{-0.07}$
& $0.56^{+0.21}_{-0.21}$ & $1.93^{+0.08}_{-0.08}$
& $0.97^{+0.37}_{-0.37}$ & $1.84^{+0.19}_{-0.19}$
\\[0.08cm]
0.90
& --- & --- 
& $0.30^{+0.14}_{-0.14}$ & $1.99^{+0.16}_{-0.16}$
& $0.55^{+0.19}_{-0.19}$ & $2.08^{+0.13}_{-0.13}$
& $0.44^{+0.30}_{-0.30}$ & $1.96^{+0.24}_{-0.24}$
\\[0.08cm]
1.10
& --- & --- 
& --- & --- 
& $0.30^{+0.12}_{-0.12}$ & $1.87^{+0.16}_{-0.16}$
& $0.39^{+0.19}_{-0.19}$ & $1.94^{+0.20}_{-0.20}$
\\[0.08cm]
1.35
& --- & --- 
& --- & --- 
& $0.23^{+0.09}_{-0.09}$ & $2.00^{+0.14}_{-0.14}$
& $0.76^{+0.43}_{-0.43}$ & $2.03^{+0.34}_{-0.34}$
\\[0.08cm]
1.75
& --- & --- 
& --- & --- 
& --- & --- 
& $0.39^{+0.22}_{-0.22}$ & $2.10^{+0.13}_{-0.13}$
\\[0.08cm]
\hline
\hline
\multicolumn{9}{l}{Star-forming} \\
0.50
& $0.09^{+0.02}_{-0.02}$ & $1.80^{+0.07}_{-0.07}$
& $0.17^{+0.05}_{-0.05}$ & $1.80^{+0.09}_{-0.09}$
& $0.27^{+0.14}_{-0.14}$ & $1.80^{+0.16}_{-0.16}$
& --- & --- 
\\[0.08cm]
0.70
& $0.09^{+0.01}_{-0.01}$ & $1.75^{+0.05}_{-0.05}$
& $0.14^{+0.02}_{-0.02}$ & $1.76^{+0.06}_{-0.06}$
& $0.15^{+0.04}_{-0.04}$ & $1.80^{+0.00}_{-0.00}$
& --- & --- 
\\[0.08cm]
0.90
& --- & --- 
& $0.11^{+0.01}_{-0.01}$ & $1.86^{+0.09}_{-0.09}$
& $0.11^{+0.01}_{-0.01}$ & $1.95^{+0.17}_{-0.17}$
& $0.54^{+0.20}_{-0.20}$ & $1.99^{+0.09}_{-0.09}$
\\[0.08cm]
1.10
& --- & --- 
& $0.12^{+0.02}_{-0.02}$ & $1.89^{+0.04}_{-0.04}$
& $0.16^{+0.02}_{-0.02}$ & $1.69^{+0.00}_{-0.00}$
& $0.39^{+0.19}_{-0.19}$ & $1.77^{+0.17}_{-0.17}$
\\[0.08cm]
1.35
& --- & --- 
& --- & --- 
& $0.15^{+0.03}_{-0.03}$ & $1.87^{+0.13}_{-0.13}$
& $0.32^{+0.14}_{-0.14}$ & $1.85^{+0.15}_{-0.15}$
\\[0.08cm]
1.75
& --- & --- 
& --- & --- 
& --- & --- 
& $0.16^{+0.03}_{-0.03}$ & $1.85^{+0.00}_{-0.00}$
\\[0.08cm]
\hline
\hline                  
\end{tabular}
\end{table*}

\begin{table*}
\caption{Fit parameters, $r_0$ (comoving and in units of $h^{-1}\mbox{Mpc}$) and $\gamma$, for the correlation functions of all galaxies as a function of mass and redshift.}
\label{tab:r0gam}
\centering          
\begin{tabular}{lcccccccc}     
\hline\hline       
$z$ & \multicolumn{2}{c}{$10^{8.6}<M[\hinvm]\leq10^{9.6}$} & \multicolumn{2}{c}{$10^{9.6}<M[\hinvm]\leq10^{10.6}$}& \multicolumn{2}{c}{$M[\hinvm]>10^{10.6}$} & \multicolumn{2}{c}{$M[\hinvm]>10^{11}$} \\
         & $r_0$ & $\gamma$& $r_0$ & $\gamma$& $r_0$ & $\gamma$& $r_0$ & $\gamma$\\
\hline
\multicolumn{9}{l}{Passive$+$Star-forming} \\
0.50
& $  3.3\pm 0.3$ & $  1.9\pm  0.3$
& $  4.5\pm 0.6$ & $  1.8\pm  0.5$
& $  4.9\pm 1.3$ & $  1.9\pm  0.9$
& --- & --- 
\\[0.08cm]
0.70
& $  3.8\pm 0.4$ & $  1.8\pm  0.3$
& $  5.0\pm 0.4$ & $  1.7\pm  0.2$
& $  6.8\pm 0.6$ & $  1.9\pm  0.3$
& $  9.4\pm 1.7$ & $  1.7\pm  0.5$
\\[0.08cm]
0.90
& --- & --- 
& $  4.0\pm 0.4$ & $  1.8\pm  0.4$
& $  5.0\pm 0.6$ & $  1.8\pm  0.4$
& $  5.8\pm 0.9$ & $  1.9\pm  0.6$
\\[0.08cm]
1.10
& --- & --- 
& $  3.9\pm 0.3$ & $  1.9\pm  0.3$
& $  5.0\pm 0.3$ & $  1.6\pm  0.1$
& $  6.9\pm 0.9$ & $  1.7\pm  0.4$
\\[0.08cm]
1.35
& --- & --- 
& --- & --- 
& $  4.9\pm 0.4$ & $  1.8\pm  0.3$
& $  8.1\pm 0.7$ & $  1.8\pm  0.3$
\\[0.08cm]
1.75
& --- & --- 
& --- & --- 
& --- & --- 
& $  6.3\pm 0.7$ & $  1.9\pm  0.4$
\\[0.08cm]
\hline
Passive  & & & & & & & &  \\
0.50
& --- & --- 
& $  5.1\pm 0.8$ & $  2.0\pm  0.6$
& $  4.7\pm 1.0$ & $  1.8\pm  0.7$
& --- & --- 
\\[0.08cm]
0.70
& --- & --- 
& $  9.8\pm 1.7$ & $  1.9\pm  0.6$
& $  7.2\pm 1.4$ & $  1.9\pm  0.7$
& $ 10.1\pm 2.1$ & $  1.9\pm  0.7$
\\[0.08cm]
0.90
& --- & --- 
& $  5.5\pm 1.3$ & $  1.9\pm  0.9$
& $  6.3\pm 1.0$ & $  2.1\pm  0.7$
& $  5.8\pm 2.0$ & $  1.9\pm  1.3$
\\[0.08cm]
1.10
& --- & --- 
& --- & --- 
& $  6.4\pm 1.4$ & $  1.9\pm  0.8$
& $  5.9\pm 1.5$ & $  1.9\pm  0.9$
\\[0.08cm]
1.35
& --- & --- 
& --- & --- 
& $  6.0\pm 1.1$ & $  2.0\pm  0.7$
& $  9.8\pm 2.7$ & $  2.0\pm  1.1$
\\[0.08cm]
1.75
& --- & --- 
& --- & --- 
& --- & --- 
& $  5.9\pm 1.6$ & $  2.1\pm  1.2$
\\[0.08cm]
\hline
\multicolumn{9}{l}{Star-forming} \\
0.50
& $  3.0\pm 0.4$ & $  1.8\pm  0.4$
& $  4.3\pm 0.7$ & $  1.8\pm  0.6$
& $  5.6\pm 1.6$ & $  1.8\pm  0.9$
& --- & --- 
\\[0.08cm]
0.70
& $  3.3\pm 0.2$ & $  1.7\pm  0.1$
& $  4.1\pm 0.3$ & $  1.8\pm  0.2$
& $  4.1\pm 0.6$ & $  1.8\pm  0.4$
& --- & --- 
\\[0.08cm]
0.90
& --- & --- 
& $  3.8\pm 0.3$ & $  1.8\pm  0.2$
& $  3.4\pm 0.2$ & $  1.8\pm  0.2$
& $  6.4\pm 1.2$ & $  1.8\pm  0.7$
\\[0.08cm]
1.10
& --- & --- 
& $  3.7\pm 0.3$ & $  1.8\pm  0.2$
& $  4.9\pm 0.3$ & $  1.6\pm  0.2$
& $  5.3\pm 1.4$ & $  1.8\pm  0.9$
\\[0.08cm]
1.35
& --- & --- 
& --- & --- 
& $  4.7\pm 0.5$ & $  1.9\pm  0.3$
& $  6.8\pm 1.6$ & $  1.9\pm  0.8$
\\[0.08cm]
1.75
& --- & --- 
& --- & --- 
& --- & --- 
& $  7.2\pm 0.8$ & $  1.9\pm  0.4$
\\[0.08cm]
\hline                  
\end{tabular}
\end{table*}

\end{document}